%% file: main.tex
\documentclass[sigconf,screen]{acmart}

\usepackage{algorithmic}
\usepackage{graphicx}
\usepackage{textcomp}
\usepackage{xcolor}
\usepackage{hyperref}
\usepackage{xspace}
\usepackage{subcaption}
\usepackage{float}
\newfloat{Listing}{tb}{lop}
\usepackage{makecell}
\usepackage{enumitem}
\usepackage{multirow}

\copyrightyear{2026}
\acmYear{2026}
\setcopyright{cc}
\setcctype{by-nc-nd}
\acmConference[ICSE '26]{2026 IEEE/ACM 48th International Conference on Software Engineering}{April 12--18, 2026}{Rio de Janeiro, Brazil}
\acmBooktitle{2026 IEEE/ACM 48th International Conference on Software Engineering (ICSE '26), April 12--18, 2026, Rio de Janeiro, Brazil}
\acmPrice{}
\acmDOI{10.1145/3744916.3787762}
\acmISBN{979-8-4007-2025-3/2026/04}



\begin{document}

\newcommand{\one}{({\em i}\/)\xspace}
\newcommand{\two}{({\em ii}\/)\xspace}
\newcommand{\three}{({\em iii}\/)\xspace}
\newcommand{\four}{({\em iv}\/)\xspace}
\newcommand{\five}{({\em v}\/)\xspace}
\newcommand{\six}{({\em vi}\/)\xspace}
\newcommand{\seven}{({\em vii}\/)\xspace}
\newcommand{\eight}{({\em viii}\/)\xspace}
\newcommand{\nine}{({\em ix}\/)\xspace}

\newcommand{\revision}[1]{#1}

\newcommand{\framework}{\textsc{NotDec}}

\title{{\framework}: WebAssembly Decompilation With Inter-Procedural Type Recovery }

\author{Jikai Wang}
\orcid{0009-0001-0887-5560}
\affiliation{%
  \institution{Huazhong University of Science and Technology}
  \city{Wuhan}
  \country{China}
}
\email{warrenwjk@gmail.com}
\authornotemark[2]

\author{Ningyu He}
\orcid{0000-0002-9980-7298}
\affiliation{%
  \institution{The Hong Kong Polytechnic University}
  \city{Hong Kong SAR}
  \country{China}
}
\email{ningyu.he@polyu.edu.hk}
\authornotemark[1]

\author{Tianming Liu}
\orcid{0000-0002-5216-933X}
\affiliation{%
  \institution{Huazhong University of Science and Technology}
  \city{Wuhan}
  \country{China}
}
\email{tmliu@hust.edu.cn}
\authornotemark[2]

\author{Junhai Wang}
\orcid{0009-0003-4328-5413}
\affiliation{%
  \institution{Huazhong University of Science and Technology}
  \city{Wuhan}
  \country{China}
}
\email{junhaiwang@hust.edu.cn}
\authornotemark[2]

\author{Haoyu Wang}
\orcid{0000-0003-1100-8633}
\affiliation{%
  \institution{Huazhong University of Science and Technology}
  \city{Wuhan}
  \country{China}
}
\email{haoyuwang@hust.edu.cn}
\authornote{Ningyu He and Haoyu Wang are the corresponding authors.}
\authornote{The full name of the authors' affiliation is Hubei Key Laboratory of Distributed System Security, Hubei Engineering Research Center on Big Data Security, School of Cyber Science and Engineering, Huazhong University of Science and Technology.}

\begin{abstract}
With WebAssembly widely supported in browsers, containers, IoT devices, and serverless platforms and increasingly adopted as a universal low‑level bytecode standard, auditing its hidden vulnerabilities and malicious intentions has become critical. 
Decompiling existing WebAssembly modules can help security researchers and end users understand binary behavior, but current tools suffer from verbose result, poor readability, and limited type recovery.

We present {\framework}, an advanced WebAssembly decompilation framework. {\framework} extends the WebAssembly type checking algorithm to lift bytecode into an SSA‑based IR, applies the inter-procedural type recovery algorithm Retypd with pointer and numeric value differentiation methods to recover complex data structures, and leverages Memory SSA alongside semantics‑preserving structured control‑flow analysis to emit readable, semantically consistent C code. 

\revision{
{\framework} achieves 100\% recompilation success rate on all 5,241 Juliet samples and all Howard dataset programs, significantly outperforming baselines including Ghidra (45.95\% success rate). On type recovery accuracy, {\framework} recovers 85.33\% of struct member accesses in real-world programs, vastly exceeding Ghidra's 9.24\%. While the full inter-procedural version faces scalability challenges on large binaries, the intra-procedural variant ${\framework}_F$ demonstrates superior efficiency, consuming less than half of Ghidra's memory and up to 97\% less execution time on unoptimized binaries.
}

\end{abstract}

\maketitle

\input{intro.tex}

\input{background.tex}

\input{method.tex}

\input{evaluation.tex}

\input{discussion.tex}

\bibliographystyle{ACM-Reference-Format}
\bibliography{ref}

\end{document}

%% file: intro.tex
\section{Introduction}
WebAssembly is a fast and compact low-level bytecode standard designed to boost web performance. It serves as a compilation target for languages such as C/C++, Rust, and Go~\cite{bringing}. As of March 2024, 99\% of monitored web browsers support it. Beyond the web, WebAssembly is also widely used in containers, IoT, and serverless platforms, offering security isolation and performance benefits~\cite{wasmnonweb}.

However, WebAssembly code still faces serious security risks. Studies show that memory safety flaws from source languages (e.g., C) persist after conversion to WebAssembly~\cite{portrisk} and can be exploited by attackers~\cite{everythingold}. Moreover, over 50\% of websites using WebAssembly deploy it for malicious activities like cryptocurrency mining~\cite{newkid}.

To address security risks from WebAssembly, researchers and users often need to audit existing modules. 
However, the source of WebAssembly code may be untrustworthy and is often not visible to clients.
While some program analysis methods, such as symbolic execution~\cite{eunomia}, can partially recover semantics, they have a high usage barrier and are limited by issues like path explosion.
Therefore, decompilation offers a more practical solution, allowing typical users to understand and analyze the internal behavior of WebAssembly modules.

Decompiling WebAssembly binaries is challenging. 
First, WebAssembly's operand stack's design complicates static code analysis, and its strict pre-execution type-checking mechanism limits code modification, necessitating conversion to a more analyzable intermediate representation (IR) to facilitate subsequent analysis and transformation.
Second, to produce readable output, recovering high-level types such as custom structs from low-level code remains difficult. Because it requires inferring types from pointer usage patterns of untyped linear memory, a task that remains a known limitation of current WebAssembly decompilers.
Finally, to produce readable C code enriched with meaningful high-level types, it is essential to carefully preserve semantic equivalence throughout the code transformation.

To address WebAssembly's operand stack and type checking challenges, we extend the type checking algorithm~\cite{wasmdocvalidation} to convert WebAssembly into an SSA-based IR, enabling the use of existing mature code analysis and optimization methods. To address the challenges of type recovery in linear memory, we introduce a state-of-the-art native binary type recovery algorithm Retypd~\cite{retypd}. 
To bridge the gap in applying type recovery algorithms, we propose the PNDiff Graph, which distinguishes pointer values from numeric values, producing precise pointer constraints for type recovery.
To maintain semantic consistency during C code generation, we employ Memory SSA~\cite{memoryssa} to ensure memory reference expressions match actual memory contents. and use a semantics-preserving structuring algorithm to convert the CFG into control flow statements, preserving result semantics.

\textbf{This work.}
In this work, we propose {\framework}, a WebAssembly decompiler that supports custom structure recovery. It comprises three modules. The WebAssembly Code Lifting module parses WebAssembly code and converts it into SSA-based IR. The Optimization and Type Recovery module optimizes the code and generates a mapping from values in the IR to high-level C language types. The Control Flow Structuring and C Code Generation module transforms the intermediate representation with high-level type information into C code.

In our evaluation using the Juliet dataset (5,241 samples) and the Howard dataset (five large-scale programs), we demonstrate both efficiency and effectiveness. For efficiency (\S\ref{ssec:rq1}), while the full inter-procedural {\framework} faces performance issues on large binaries (e.g., timing out in some cases), its intra-procedural variant ${\framework}_{F}$ achieves less time and memory consumption compared to Ghidra, particularly on unoptimized binaries. For effectiveness (\S\ref{ssec:rq2}), our approach excels in type recovery, successfully reconstructing 85.33\% of struct member accesses in the Howard dataset, vastly outperforming Ghidra, which only recovers 9.24\%.
An ablation study (\S\ref{eva:abla}) shows that code optimization combined with low-level pattern matching effectively reduces redundancy in binary code. The type recovery module contributes to the recovery of all of struct types and their associated accesses. Furthermore, deconstructing the stack into concrete variables yields additional reduction in LoC.

\textbf{Contribution.}
In this work, our major contributions are:

\begin{itemize}[leftmargin=*]
 \item \textbf{WebAssembly Decompiler.} We present a novel WebAssembly decompiler that supports recovery of user-defined (custom) structure types and produces re-compilable high-level code.
 \item \revision{\textbf{Decompilation System Design.} We integrate mature compiler optimization techniques into decompilation and combine them with type-recovery algorithms (\S\ref{mid:lowpat}). Furthermore, we model each function's stack in linear memory as a single large structure and leverage compiler optimizations to decompose it into individual stack variables (\S\ref{stacksroa}). Besides, we introduce Memory SSA to resolve whether memory reads have been updated when transforming SSA-based intermediate representations back into non-SSA code (\S\ref{method:cgen}).}
 \item \textbf{Efficiency.} {\framework} achieves 15.9× faster execution and 4.8× lower memory consumption than Ghidra on the Juliet dataset, with its intra-procedural variant ${\framework}_F$ scales effectively to large binaries, consuming less than half of Ghidra's memory and up to 97\% less execution time on unoptimized code.
 \item \textbf{Effectiveness.} {\framework} is the only tool achieving 100\% recompilation success across all evaluated binaries and recovers 85.33\% of struct member accesses compared to Ghidra's 9.24\%, demonstrating substantial improvements in code readability and semantic reconstruction.
 \item \textbf{Open Source}: {\framework} is available at \url{https://github.com/NotDec/NotDec}.
 \end{itemize}

%% file: background.tex
\section{Background \& Challenges}

\subsection{Background}

\noindent
\textbf{WebAssembly}.
WebAssembly (Wasm) was initially created to improve web application performance through its lightweight and compact binary format. Its loading, validation, and execution speeds far exceed those of traditional JavaScript, while running in a secure sandbox and working in conjunction with JavaScript~\cite{bringing}. 
Today, WebAssembly has evolved into a general-purpose low-level bytecode standard that supports various programming languages such as C/C++, Rust, and Go~\cite{wasmbench}. 
To abstract various memory management methods of different languages, WebAssembly retains a naive array-like linear memory structure.

\noindent
\textbf{Structured Control Flow \& Type Checking}. 
WebAssembly features unique structured control flow and type-checking mechanisms. 
Unlike traditional assembly code that allows arbitrary jumps, WebAssembly uses \texttt{block}, \texttt{loop}, and \texttt{end} to explicitly mark the nested levels of the code. Control flow jumps are only permitted from within a nested block to the end of the block (for \texttt{block} blocks) or to the beginning (for \texttt{loop} blocks)~\cite{wasmspec}. 

WebAssembly has an operand stack from which all instructions read operands and push the results onto the stack. There is also a type checking algorithm~\cite{wasmdocvalidation} that ensures that the values obtained from the operand stack are always of the correct type. 
Additionally, each block declares parameter and return types. At the beginning and the end of each block, if the types the remained operands in the stack mismatch with the parameter or the return type, the type check will fail.
To this end, the WebAssembly module will be rejected during loading, thereby preventing the code from running.

\noindent
\textbf{WebAssembly Decompilation.}
Existing academic WebAssembly decompilers have limited capabilities and often produce code lacking high-level types.
WaDec~\cite{wadec} converts WebAssembly bytecode to C code using fine-tuned large language models, but it suffers from hallucinations and cannot ensure semantic consistency or even syntactic correctness.
SnowWhite~\cite{snowwhite} infers high-level types for parameters and return values based on sequence generation models, yet it only recognizes common data structure patterns (e.g., FILE descriptors) and cannot handle custom ones.
We also underline that both learning-based methods also lack interpretability.

As for industrial tools, IDA's Hex-rays decompiler~\cite{idadec} supports loading WebAssembly but does not support decompilation. Ghidra~\cite{ghidra} can support WebAssembly decompilation by installing a community plugin (ghidra-wasm-plugin). 
There are also some decompilation tools that merely generate ``low-level'' C code, such as wasm2c from WABT (The WebAssembly Binary Toolkit) and wasmdec\footnote{https://github.com/wwwg/wasmdec} (the highest-starred "decompiler" on GitHub). Their output is typically hard to read, characterized by excessive \texttt{goto} statements and pointer operations expressed as numeric calculations—only cast to pointer types during memory accesses. 
These limitations significantly reduce code readability and hinder further analysis.

\subsection{Challenge \& Solution}
There are three main challenges in decompiling WebAssembly code. 

\noindent
\textbf{Challenge \#1: Handling the operand stack of WebAssembly for better code analysis and optimization}. 
The operand stack obscures use-def relationships between instructions, complicating code analysis. Additionally, WebAssembly's type-checking mechanism restricts stack usage, making code transformations difficult~\cite{wasmdocvalidation}. For example, if instructions are inserted without considering stack balance, it can easily lead to misalignment of values on the operand stack, causing the code to fail type checking and thus become invalid~\cite{brewasm}. Effective analysis and transformation are essential for decompilation, posing a significant challenge.

\noindent
\textbf{Challenge \#2: Recovering local variables and custom data structures.}
WebAssembly's operand stack cannot take addresses or support aggregate types, so compilers often maintain another call stack in the untyped linear memory. This results in a significant loss of type information for local variables, requiring their high-level types to be inferred from stack memory access patterns, similar to traditional binary reverse engineering. Moreover, the code may contain complex custom structures type (such as recursive types), and the usage of types may be distributed across different functions.

\noindent
\textbf{Challenge \#3: Reuse memory load expression to make code concise while ensuring semantic consistency.}
If a memory value is modified between its load and subsequent use, the load no longer corresponds to the current variable value but to its prior state, necessitating a temporary variable. Ensuring that load expressions map to the correct value, especially when they may represent variable accesses, is a key challenge.

\noindent
\textbf{Our Solution}. 
For \textbf{Challenge \#1}, we lift WebAssembly bytecode to an SSA-based IR, allowing existing compilation optimizations and analyses to be applied (\S~\ref{method:lift}). Leveraging the static determinism of the operand stack, we translate instructions directly into a register-based SSA form, preserving original semantics.

For \textbf{Challenge \#2}, we applied the state-of-the-art binary type recovery algorithm, Retypd~\cite{retypd} (\S~\ref{mid:retypd}). Additionally, to distinguish between pointer and numeric types to extract more constraints on types, we proposed PNDiff graph and apply a set of inference rules.

For \textbf{Challenge \#3}, we employ Memory SSA to determine if a memory location remains unchanged after its last load. A temporary variable is introduced only when a potential modification is detected, thereby balancing correctness with readability (§\ref{method:cgen}).

%% file: method.tex
\section{Design Overview}

\begin{figure*}[t]
    \centering
    \includegraphics[width=0.8\linewidth]{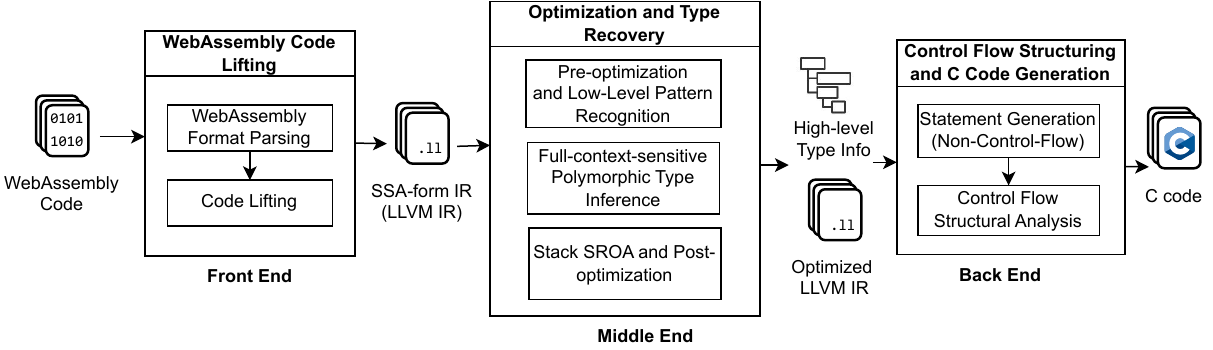}
    \caption{The overview of our WebAssembly decompilation method.}
    \label{overview}
    \vspace{-0.1in}
    \Description[Our method have 3 parts.]{Our method consists of 3 parts, WebAssembly code lifting, optimization and type recovery, and control flow structuring and C code generation.}
\end{figure*}

The architecture and workflow of our proposed decompilation process is shown in \autoref{overview}. 
Similar to a compiler's architecture, our approach is divided into three parts: 
\one \textit{Front End.} The front end takes a WebAssembly code file as the input, in which the WebAssembly code lifting module is responsible for converting it into an SSA-based intermediate representation (\textit{e.g.,} LLVM IR);
\two \textit{Middle End.} The middle end is mainly responsible for optimization and type recovery, outputting the optimized IR and high-level type information for each value in the IR; and
\three \textit{Back End.} In the back end, the control flow structuring and C code generation module is responsible for converting the IR with high-level type information into a more readable piece of C code, which will be output finally.


\subsection{Front End: Code Lifting} 
\label{method:lift}
As noted in \textbf{Challenge \#1}, mandatory type verification restricts WebAssembly programs, making direct transformation and optimization on WebAssembly instructions difficult.
To facilitate better code analysis and transformation, we convert WebAssembly code into a register-based Static Single Assignment (SSA) Intermediate Representation (IR) format. The conversion to SSA is also mentioned in the design rationale document~\cite{wasmrationale}, and implemented by V8 and SpiderMonkey~\cite{bringing}.
We underline that SSA is widely used in compiler optimizations and in decompilers (like Ghidra's P-Code IR~\cite{ssa4dec}). Converting to SSA enables the application of a wide range of existing and mature code analysis and optimization algorithms.

We first build use-def relationships for all instructions by extending the specification’s type checking algorithm~\cite{wasmdocvalidation} described by the specification, transforming the operand-stack-based form into a register-based SSA form.
Then, we continue the process according to instruction types:
\begin{itemize}[leftmargin=*]
\item \textbf{Variable Instructions.} For WebAssembly's global and local variables of simple types, \textit{i.e.,} i32, i64, f32, and f64. We allocate mutable global or local memory. Instructions like \texttt{local.get} and \texttt{global.set} become loads and stores to that memory.
\item \textbf{Memory Instructions.} We map linear memory to a global byte array of equal size and transform all memory access operations like \textit{i32.load} into accesses to this array. This memory variable will later be split during type analysis (see \S~\ref{stacksroa}).
\item \textbf{Control Instructions.} Structured control flow is converted to a standard control flow graph. For \texttt{br\_table}, we resolve all possible targets and emit a \texttt{switch} instruction in the IR.
\item \textbf{Table/Element Section \& Indirect Call.} In WebAssembly, indirect call locates the target function via a table (defined in the element section) storing function references. We create a corresponding global function pointer array for each function table in the element section. When encountering a \texttt{call\_indirect}, we access this global array by index, retrieve the function pointer, and then generate a function pointer call.
\item \textbf{Vector Instructions.} We create corresponding SIMD instructions in the IR according to its semantics.
\end{itemize}

\subsection{Middle End: Optimization \& Type Recovery} \label{method:tr}

\begin{figure}[t]
    \centering
    \includegraphics[width=0.9\linewidth]{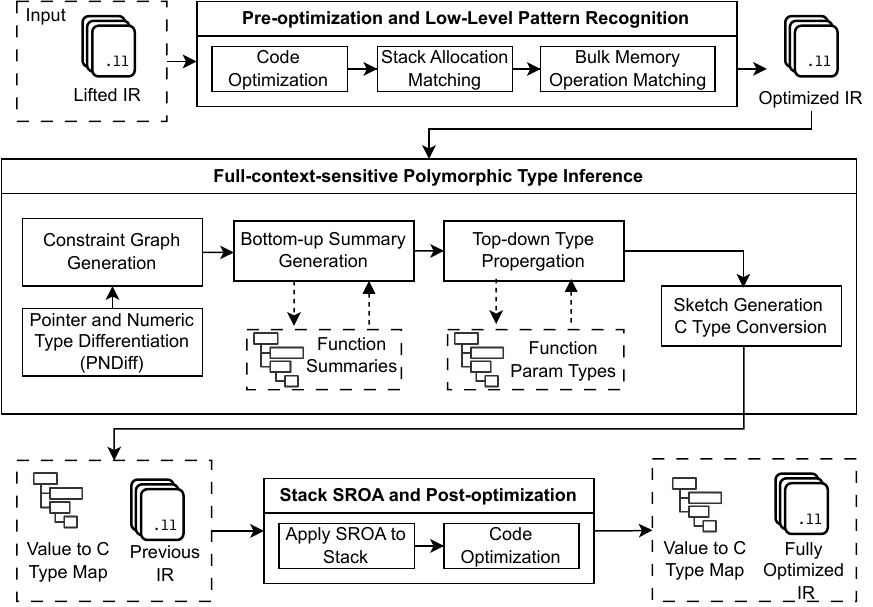}
    \caption{The main process of type recovery.}
    \label{typeinfer}
    \vspace{-0.1in}
    \Description[The main process of type recovery.]{type recovery consists of 3 parts, Pre-optimization and Low-Level Pattern Recognition, Full-context-sensitive Polymorphic Type Inference and Stack SROA and Post-optimization.}
\end{figure}

As shown in \autoref{overview}, the middle end takes the lifted IR file as input, performs optimization and type recovery, and outputs an optimized IR along with a mapping from IR values to high-level types.
The detailed architecture (\autoref{typeinfer}) involves three main steps: \textit{pre-optimization and low-level pattern recognition}, \textit{full-context-sensitive polymorphic type inference}, and \textit{stack SROA and post-optimization}.

\subsubsection{Pre-optimization and Low-level Pattern Recognition} \label{mid:lowpat}

\begin{figure}[t]
    \centering
    \includegraphics[width=0.95\linewidth]{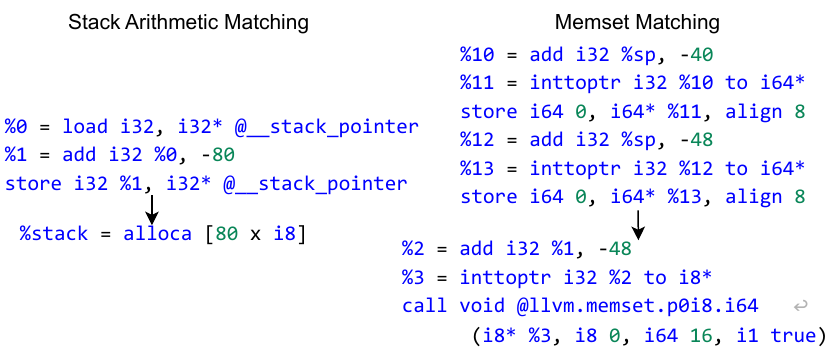}
    \caption{Examples shows the conversion of instructions for stack allocation matching and memset matching.}
    \label{stackexp}
    \vspace{-0.1in}
    \Description[Examples shows the conversion of instructions for stack allocation matching and memset matching.]{Examples shows the conversion of instructions for stack allocation matching and memset matching.}
\end{figure}

We first apply code optimization to eliminate redundancies in the lifted code (especially from unoptimized binaries).
Additionally, reducing the size of the code can increase the efficiency of subsequent analyses. 
As for the pattern recognition, it focuses on two key tasks: identifying stack allocations and recognizing common memory operations such as \texttt{memcpy} and \texttt{memset}. 
For stack allocation, we replace low-level stack pointer manipulations with explicit stack space allocations, making stack memory visible to type analysis and enabling correct type inference for local variables.
As for \texttt{memcpy} and \texttt{memset}, compilers often translate them in WebAssembly into sequences of fixed-size (e.g., 8-byte) load and store instructions. If left unhandled, these patterns can mislead type recovery by suggesting incorrect data layouts or member structures. Recognizing and abstracting them back into high-level operations helps preserve accurate type information.

These steps are implemented as a sequence of transformation passes over the intermediate representation (IR).
We perform pattern recognition after code optimization because compiler optimizations often rewrite related operations into standardized,  \textit{i.e.,} \textit{canonical}, forms. This simplifies pattern matching by reducing variations we need to handle.
For stack pointer operations, we look not only for the typical stack pointer decrement at the beginning of a function (the entry block) but also for dynamic stack allocations that occur elsewhere in the code—such as those generated by C’s \texttt{alloca} function.
For \texttt{memset} and \texttt{memcpy}, we scan for sequences of consecutive memory stores to the same region, and replace them with a single intrinsic \texttt{memset} or \texttt{memcpy} call. For example, as shown on the left side of \autoref{stackexp}, three instructions identified as adjusting the stack pointer are transformed into a single, clear stack allocation statement. On the right, a sequence of consecutive store instructions are collapsed into a single \texttt{memset} call.

\begin{figure}[t]
    \centering
    \begin{subfigure}{0.48\linewidth} 
        \includegraphics[width=\linewidth]{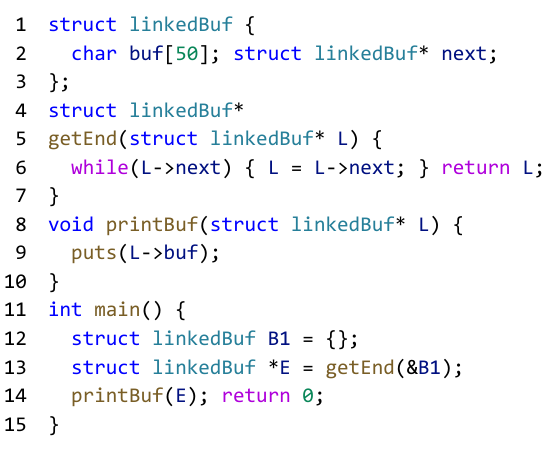}
        \caption{Original code to be compiled to WebAssembly.}
        \label{motivatingorig}
    \end{subfigure}
    \hfill
    \begin{subfigure}{0.5\linewidth}
        \includegraphics[width=\linewidth]{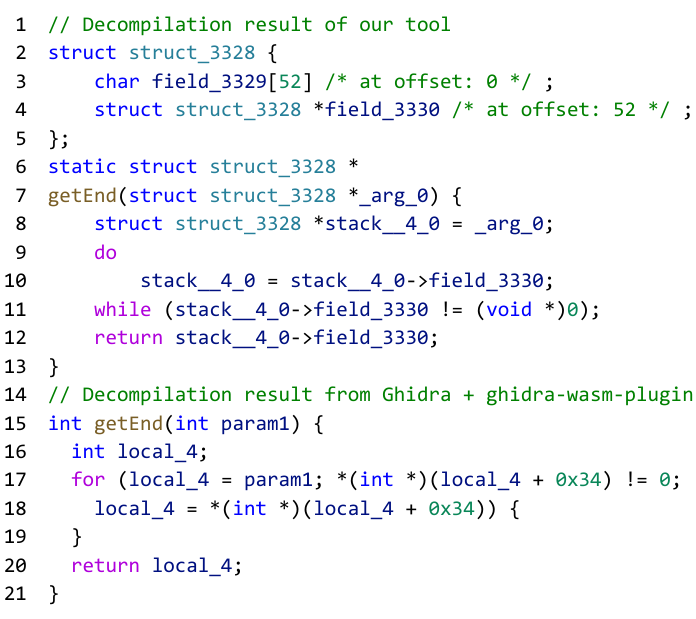}
        \caption{Decompilation result of \texttt{getEnd} using our tool and Ghidra.}
        \label{motivatingdec}
    \end{subfigure}
    \vspace{-0.1in}
    \caption{Sample code that requires inter-procedural type analysis.}
    \label{motivating}
\end{figure}

\subsubsection{Full-context-sensitive Polymorphic Type Inference} \label{mid:retypd}

To improve code readability, we need to analyze the high-level type information within the code. The input for this stage is the optimized IR, and after analysis, the output will be a mapping of each value in the IR to its corresponding high-level C type.

Due to the fact that WebAssembly's linear memory is untyped and byte-addressable, its type recovery closely resembles that of native binary code. 
Therefore, we attempt to introduce a state-of-the-art binary type recovery algorithm, Retypd~\cite{retypd}, into WebAssembly. 
For example, when decompiling the WebAssembly code generated from the code in \autoref{motivatingorig}, our tool can produce results for this sample function with the correct type information (in \autoref{motivatingdec}, line 2-13). In contrast, the code generated by Ghidra contains pointer arithmetic and type casting (at line 17-18), impacting the readability.

\paragraph{Pointer and Numeric Type Differentiation (PNDiff)}
Retypd cannot be directly applied as it relies on correctly identifying, among integer-typed values, those that serve as memory addresses (i.e., pointers) and their associated pointer arithmetic operations, which are essential constraints for subsequent type inference. This differentiation is critical because: \one pointer addition typically changes the pointer’s target type; \two constant pointer offsets often indicate accesses to structure members; and \three variable pointer offsets may suggest array indexing.

\begin{table}[t]
\centering
\caption{Inference rules for addition and subtraction.}
\resizebox{0.95\linewidth}{!}{
\begin{tabular}{c|l|l}
\toprule
\textbf{Rule} & \boldmath$X+Y=Z$ & \boldmath$X-Y=Z$ \\ \midrule
1 & $X\colon\mathrm{Num},\,Y\colon\mathrm{Num}\;\to\;Z\colon\mathrm{Num}$ 
  & $X\colon\mathrm{Num}\;\to\;Y\colon\mathrm{Num},\,Z\colon\mathrm{Num}$ \\

2 & $Z\colon\mathrm{Num}\;\to\;X\colon\mathrm{Num},\,Y\colon\mathrm{Num}$ 
  & $Y\colon\mathrm{Num},\,Z\colon\mathrm{Num}\;\to\;X\colon\mathrm{Num}$ \\

3 & $X\colon\mathrm{Ptr}\;\to\;Y\colon\mathrm{Num},\,Z\colon\mathrm{Ptr}$ 
  & $Y\colon\mathrm{Num},\,Z\colon\mathrm{Ptr}\;\to\;X\colon\mathrm{Ptr}$ \\

4 & $Y\colon\mathrm{Num},\,Z\colon\mathrm{Ptr}\;\to\;X\colon\mathrm{Ptr}$ 
  & $Y\colon\mathrm{Ptr}\;\to\;X\colon\mathrm{Ptr},\,Z\colon\mathrm{Num}$ \\

5 & $Y\colon\mathrm{Ptr}\;\to\;X\colon\mathrm{Num},\,Z\colon\mathrm{Ptr}$ 
  & $X\colon\mathrm{Ptr},\,Z\colon\mathrm{Num}\;\to\;Y\colon\mathrm{Ptr}$ \\

6 & $X\colon\mathrm{Num},\,Z\colon\mathrm{Ptr}\;\to\;Y\colon\mathrm{Ptr}$ 
  & $X\colon\mathrm{Ptr},\,Y\colon\mathrm{Num}\;\to\;Z\colon\mathrm{Ptr}$ \\

7 & — 
  & $X\colon\mathrm{Ptr},\,Z\colon\mathrm{Ptr}\;\to\;Y\colon\mathrm{Num}$ \\ \bottomrule
\end{tabular}
}
\label{tab:pnirule1}
\end{table}

This problem is a fundamentally undecidable challenge in native binary analysis. WebAssembly faces the same issue because its memory instructions also use integer types (e.g., \texttt{i32}), making pointers syntactically indistinguishable from integers.
Therefore, we introduce \textit{PNDiff graph}, a graph-based solution that adapts established principles from native binary analysis for differentiating pointers from integers (i.e., rules in \autoref{tab:pnirule1}).

Specifically, the PNDiff analysis proceeds as follows:
\begin{enumerate}[leftmargin=*]
\item Node creation. Each program value is represented as a node in the PNDiff graph, annotated with a \textit{PN Type}, which is one of four categories: \texttt{Ptr} (pointer), \texttt{Num} (numeric), \texttt{Unknown} (could be either), or \texttt{Non-PN} (e.g., floating-point values).
\item Initialization. We seed the graph with known type information. For addresses used in load/store instructions are initialized as \texttt{Ptr}. Values involved in numeric-only operations (e.g., multiplication, division) are marked as \texttt{Num}.
\item Node merging. Dataflows like assignment operations induce equivalence between source and destination values, so their corresponding nodes are merged. Similarly, equality comparisons (\texttt{==}, \texttt{!=}) imply that both operands must be of the same category (both pointers or both numbers), prompting node merging.
\item Constraint collection. For every addition or subtraction instruction, we record a constraint involving its three operands (two inputs and one result) as a triple of PN nodes.
\item Rule-based inference. We iteratively apply inference rules (summarized in \autoref{tab:pnirule1}) to propagate type information across constraints. For instance, Rule 7 for subtraction states that if both the minuend and difference are pointers, the subtrahend must be numeric—reflecting the semantics of pointer subtraction.
\end{enumerate}

\begin{table}[t]
  \centering
  \caption{Additional inference rules for addition and subtraction using equality relation.}
  \resizebox{0.95\linewidth}{!}{
  \begin{tabular}{c|l|l}
    \toprule
    \textbf{Rule} & \boldmath$X+Y = Z$ & \boldmath$X - Y = Z$ \\
    \midrule
    1 & $X = Y\;\to\;X:\mathrm{Num},\,Y:\mathrm{Num}$ 
      & $Y = Z\;\to\;X:\mathrm{Num},\,Y:\mathrm{Num}$ \\
    2 & $X = Z\;\;\to\;\;Y:\mathrm{Num}$ 
      & $X = Y\;\;\to\;\;Z:\mathrm{Num}$ \\
    3 & $Y = Z\;\;\to\;\;X:\mathrm{Num}$ 
      & $X = Z\;\;\to\;\;Y:\mathrm{Num}$ \\
    \bottomrule
  \end{tabular}
  }
  \label{tab:pnirule2}
\end{table}

Due to node merging in step (iii), multiple syntactic values may map to the same PN node, establishing an equivalence relation over types. We exploit this to derive additional inference rules based on type equality, as shown in Table~\ref{tab:pnirule2}. For example, If $X=Y$ in an addition $X+Y=Z$, then both must be numeric, since adding two pointers is semantically invalid.
If $X=Z$, then $Y$ must be numeric, as pointer arithmetic preserves the pointer nature of the base operand. These equality-aware rules can further enhance type resolution.

\begin{figure}[t]
    \centering
    \includegraphics[width=0.8\linewidth]{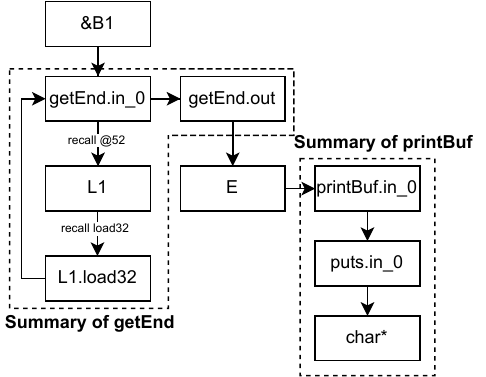}
    \caption{Constraint graph for main.}
    \label{typeexpcg}
    \vspace{-0.1in}
    \Description[Constraint graph for main.]{Constraint graph for main.}
\end{figure}

\paragraph{Constraint Graph Generation}
Generating a constraint graph is the first step of the type recovery by Retypd. Specifically, each value in the program is mapped to a node in the constraint graph.
Then we create edges according to the constraints posed by the instructions. For assignment instructions, we create subtype relationships without edge labels. For load/store instructions, we create an edge labeled as ``recall load/store'', pointing from the address value to the value in memory. For pointer addition edges, we create a ``recall offset'' edge, pointing from the pointer operand of the addition to the result of the addition.
For example, in \autoref{typeexpcg}, the constraint graph corresponding to the code in \autoref{motivatingorig} is displayed. The \texttt{getEnd.in\_0} represents the first parameter L of \texttt{getEnd}. The operation \texttt{L=L->next} in the \texttt{getEnd} function includes member access (pointer addition, ``recall edge @52''), memory loading (``recall load32 edge''), and a subtype edge.

\paragraph{Bottom-up Summary Generation \& Top-Down Type Propagation}
Similar to other summary-based full-context-sensitive static analyses \cite{fulcra}, Retypd consists of a bottom-up summary generation step and a top-down type propagation step. 
Specifically, the bottom-up summary generation involves extracting the summary for each function, which indicates the impact of the callee on the caller's type. 
Since the generation of each function's summary depends on the summaries of other functions it calls, we traverse the call graph (with SCCs collapsed) bottom-up to generate the summaries. 
For example, the simplified constraint graph of the main function in \autoref{motivatingorig} is shown in \autoref{typeexpcg}. When generating the summary for the main function, the summaries of the callees (including \texttt{getEnd} and \texttt{printBuf}) are first introduced based on the initial constraint graph. Then, the algorithm described in Retypd is used to simplify the constraint graph, which is stored as the summary of the current function.

As for the top-down type propagation, it involves finding all the types passed by its callers and merging these types to derive the most precise parameter type for each function, which is used for the final type computation. For example, in the \autoref{motivatingorig} example of the \texttt{printBuf} function, it is reasonable to consider the type of its first parameter as \texttt{char*}. However, during the top-down process, it can be inferred that all calls to \texttt{printBuf} pass in the type \texttt{linkedBuf*}, which is a subtype of \texttt{char*} according to width subtyping. Therefore, the parameter of \texttt{printBuf} is set to the more precise type \texttt{linkedBuf*}.

\revision{We underline such an inter-procedural analysis step is optional. Although it enables type information propagation across the whole program, it also dramatically increases the complexity. Its intra-procedural analysis complexity is $O(n^3)$ with respect to the number of instructions, and in the worst case, polymorphic type functions can introduce exponential complexity with the depth of calls. 
As shown in \S~\ref{ssec:rq1}, it may require more than 24 hours for a relative large program. By disabling this step and restricting the analysis scope to the intra-function level, we could achieve a balance between effectiveness and performance. In such cases, structures used inside a function can still be identified, but member hints resulting from operations outside the function are no longer taken into account.
As for the full-context-sensitivity without a call-stack depth limitation, it may lead to infinite recursion in the presence of cycles in the call graph, i.e., loops. To address this, Retypd disables context sensitivity within cycles by collapsing strongly connected components (SCCs) in the call graph into single nodes. This transformation produces a directed acyclic graph, only on which the full context-sensitive analysis is then applied.}

\paragraph{Sketches Generation and C Type Conversion} 
Based on the final constraint graph obtained from the previous step, processing through Retypd can yield graph based type representations, referred to as \textit{Sketches}.
\revision{Then, by collecting pointer offset edges and subsequent memory read/write edges, we can recover custom structure types and generate type declarations, thereby converting the Sketch graph into C types.}
\revision{In the example from \autoref{typeexpcg}, without inter-procedual type analysis, a decompiler would not know the specific types of \texttt{B1}. After including type summary from callee, we can follow the pointer offset edge ``recall @52'' and load edge ``recall load32'', and deduce that B1 has a 32-bit structure member at offset 52, thus setting \texttt{B1} as a structure pointer type.}

\subsubsection{Stack SROA and Post-optimization} \label{stacksroa}
Since the previous type inference steps do not modify the IR, the stack space allocation for functions retains the form after stack pointer operations matching, which means the entire stack space of the function is viewed as a large structure containing all local variables. To make the decompiled result closer to the original code, we employ the method of Scalar Replacement of Aggregate (SROA) from compilation optimization to split the structure into local variables. After splitting, new optimization opportunities may be exposed in the code. Therefore, we apply compiler optimizations again to simplify the code.

\subsection{Back End: C Code Generation}
\label{method:cgen}
In order to convert the SSA and control flow graph-based intermediate representation into more readable C code, we need to address \one the conversion of non-control flow instructions, and \two the structural analysis of control flow.

\textbf{Memory SSA for Correct and Readable Load Handling}.
When generating C code from our SSA‑based IR, we convert side effect instructions (e.g., stores) to statements, and we defer insertion of pure compute expressions until they are needed by other instructions, and cache these expressions in a map for later use.
A key challenge arises in keeping loads semantically consistent with subsequent updates to the same memory location. For instance, consider:

{
    \centering
    \includegraphics[width=0.85\linewidth]{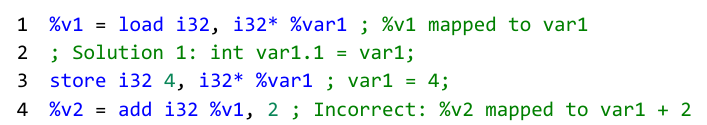} 
}

\noindent If we naively map \texttt{\%v2} as \texttt{var1+2}, we lose the fact that \texttt{\%v1} should refer to the pre‑store value of var1. A simple fix is to insert a temporary variable at the load, as shown in the second line.
Introducing a fresh temporary for every load instruction would preserve semantics but at the cost of cluttering the generated code with too many temporaries, hurting readability.

To balance correctness and readability, we integrate Memory SSA~\cite{memoryssa} to identify situations where temporary variable is not required. \revision{Memory SSA addresses our queries regarding interactions between memory operations by identifying potential pointer aliasing situations through fast intra-procedural analysis.} First, when we actually add statements or expressions to the AST, we traverse the entire AST to identify nodes corresponding to original load instructions. Second, For each such node (at current insertion point), we query Memory SSA to ask: ``Has the memory location read by this load been modified since it was last loaded?''. If not modified, then this usage is correct. If modified, we locate the original load’s insertion point, introduce a temporary local there, and rewrite the AST node to refer to that temporary variable.

\textbf{Structural Analysis of Control Flow}. Usually IR adopts a representation based on control flow graphs. To convert it into a representation in C language based on control flow statements, we need to employ decompilation structural analysis algorithms. Existing research on structure analysis is already quite mature \cite{dream,phoenix,comb,sailr}. We implemented a semantic-preserving structural analysis algorithm \cite{phoenix,sailr} to ensure that the generated code is semantically consistent with the control flow of the IR.

%% file: evaluation.tex
\section{Implementation \& Evaluation}
\label{sec:imp}

\noindent
\textbf{Implementation.} 
We implemented three main modules as separate reusable projects. 
Specifically, the front end is implemented with approximately 3K lines of C++ code, in which the WebAssembly code parsing relies on the WebAssembly Binary Toolkit (WABT) project. 
The middle end uses LLVM IR as the intermediate language and implements the low-level pattern matching and type recovery algorithms with about 9K lines of C++ code.
The back end generates C code based on the Clang AST, using approximately 6K lines of C++ code.

\noindent
\textbf{Experiment Setup.}
All experiments were conducted on a computer with 64GB of memory, two NVIDIA GeForce RTX 2080 Ti graphics cards, and an Intel i9 9900 CPU, running Ubuntu 22.04.5.

\noindent
\textbf{Benchmark.} 
\revision{Our dataset consists of the Juliet dataset~\cite{datasetsac} and Howard dataset~\cite{howard}.}
the Juliet dataset contains 5,241 samples from the Juliet Test Suite 1.3 C code vulnerability dataset, which were selected by a prior study on porting code to WebAssembly~\cite{sac22port}.
\revision{Each of these samples contains 3–6 simple functions, making them relatively small in code size.}
\revision{The Howard dataset comprises five substantial programs, \textit{i.e.,} \texttt{fortune}, \texttt{grep}, \texttt{gzip}, \texttt{lighttpd}, and \texttt{wget}, with source code sizes ranging from 2K to 46K lines of C code. We used the latest release versions of these projects as of October 2025.}

We compiled all samples into WebAssembly using wasi-sdk-20~\cite{wasisdk}, adding relevant compilation options (like \texttt{--no-standard\--libraries} and \texttt{-fno-builtin}) to avoid introducing unnecessary library functions. \revision{For the Howard dataset, some relied on features not supported by the compilation toolchain. We addressed this by introducing additional header files and avoid linking to the actual libraries to allow compilation to proceed. The resulting WebAssembly binaries, while suitable for decompilation and static analysis, cannot be actually executed. We obtained two different binaries for each project with \texttt{-O3} and \texttt{-O0} compilation options.}

\noindent \textbf{Baselines.}
To the best of our knowledge, there are only three available WebAssembly decompilation tools as follows.
\one \textbf{ghidra-wasm-plugin}~\cite{ghidrawasmplugin}: Ghidra does not natively support WebAssembly, but the community-developed ghidra-wasm-plugin enables WebAssembly decompilation in Ghidra by converting WebAssembly into the P-code intermediate representation used by Ghidra's decompiler. 
\two \textbf{WaDec}~\cite{wadec}: WaDec is a WebAssembly bytecode-to-C code solution based on fine-tuned large language models. 
\three \textbf{WasmDec}~\cite{wasmdec}: WasmDec is the most starred WebAssembly decompiler on GitHub. It simply converts WebAssembly into the ``low-level'' C code without further stack pointer handling or type recovery.

\noindent
\textbf{Evaluation.}
In this work, we aim to answer the following research questions (RQs):

\begin{itemize}[leftmargin=*]
    \item \textbf{RQ1:} How efficient is {\framework} compared with baselines?
    \item \textbf{RQ2:} How effective is {\framework} compared with baselines?
    \item \textbf{RQ3:} How important is the optimization and type recovery phase mentioned in \S~\ref{method:tr} for decompilation?
\end{itemize}

To answer RQ1, we measure success rates, runtime, and memory consumption of {\framework} and three baselines (WasmDec, Ghidra, WaDec). 
For RQ2, we evaluate effectiveness from three aspects: recompilation success, line of code (LoC), and \revision{type recovery accuracy, measured by two metrics: the number of correctly reconstructed structs, and the number of struct member accesses recovered from low-level pointer operations.}
For RQ3, we perform an ablation study with four {\framework} variants on Juliet samples, to show the importance of the optimization and type recovery phase.

\subsection{RQ1: Efficiency}
\label{ssec:rq1}

\begin{table}[t]
\centering
\caption{Statistics on 5,241 samples from Juliet C dataset.}
\resizebox{0.95\linewidth}{!}{
\begin{tabular}{@{}r|l|l|l@{}}
\toprule
Tool        & \# Success           & Avg. Time (s)              & Avg. Memory (MB) \\ \midrule
{\framework}      & 5,241     & 1.73     & 119.13 \\
Ghidra      & 5,241     & 27.62     & 572.14 \\
WaDec       & 1,940      & 108.11      & 9,890.45 \\
Wasmdec     & 5,241    & 0.02    & 6.46 \\  \bottomrule
\end{tabular}
}
\label{rq1-1}
\end{table}

\begin{table*}[t]
\centering
\caption{Average time and memory consumption on samples (\texttt{-O0} and \texttt{-O3} enabled) from Howard dataset.}
\resizebox{0.95\linewidth}{!}{

\begin{tabular}{@{}rlllllllllll@{}}
\toprule
\multirow{2}{*}{Benchmark} & \multicolumn{1}{c}{\multirow{2}{*}{\begin{tabular}[c]{@{}c@{}}\#Inst in \\ binary\end{tabular}}} & \multicolumn{5}{c}{Avg. Time (s)} & \multicolumn{5}{c}{Avg. Memroy (MB)} \\ \cmidrule(l){3-12} 
 & \multicolumn{1}{c}{} & {\framework} & \multicolumn{1}{c}{${\framework}_{F}$} & \multicolumn{1}{c}{Ghidra} & \multicolumn{1}{c}{WaDec} & \multicolumn{1}{c}{WasmDec} & \multicolumn{1}{c}{{\framework}} & \multicolumn{1}{c}{${\framework}_{F}$} & \multicolumn{1}{c}{Ghidra} & \multicolumn{1}{c}{WaDec} & \multicolumn{1}{c}{WasmDec} \\ \midrule
fortune (\texttt{O0}) & 13,653 & 4.08 & 3.25 & 23.25 & 1,431.54 & 0.01 & 193.70 & 99.84 & 1,017.85 & 10,523.44 & 7.91 \\
fortune (\texttt{O3}) & 4,797 & 2.07 & 2.38 & 12.44 & 145.83 & 0.02 & 122.29 & 101.00 & 730.70 & 10,517.71 & 7.56 \\
grep (\texttt{O0}) & 245,806 & 51,092.00 & 68.05 & 522.53 & 8,508.00 & 0.10 & 13,397.80 & 452.22 & 2,492.05 & 10,521.06 & 27.82 \\
grep (\texttt{O3}) & 93,227 & >24h & 46.81 & 45.72 & 15,897.00 & 0.04 & 4,815.32 & 510.36 & 907.94 & 10,523.42 & 15.62 \\
gzip (\texttt{O0}) & 124,327 & >24h & 39.65 & 1,665.50 & 3,490.40 & 0.05 & 6,325.64 & 247.40 & 14,244.22 & 10,527.30 & 17.08 \\
gzip (\texttt{O3}) & 51,279 & 3,657.00 & 29.35 & 55.43 & 4,230.00 & 0.03 & 2,010.41 & 297.44 & 1,735.79 & 10,523.79 & 11.12 \\
lighttpd (\texttt{O0}) & 484,209 & >24h & 141.06 & 630.11 & 31,747.00 & 0.19 & 20,617.43 & 898.55 & 7,344.68 & 10,527.17 & 48.91 \\
lighttpd (\texttt{O3}) & 141,083 & >24h & 69.89 & 55.43 & >24h & 0.06 & 3023.89 & 795.36 & 2,194.95 & 10,515.05 & 20.04 \\
wget (\texttt{O0}) & 433,082 & >24h & 129.01 & 1,550.37 & 25,069.00 & 0.17 & 8152.79 & 730.00 & 14,742.57 & 10,520.45 & 44.56 \\
wget (\texttt{O3}) & 139,729 & >24h & 78.32 & 70.22 & 42,774.00 & 0.06 & 8,645.49 & 730.28 & 2,472.37 & 10,514.49 & 19.82 \\
\bottomrule
\end{tabular}

}
\label{rq1-1-2}
\end{table*}

\begin{figure}[t]
    \centering
    \includegraphics[width=\linewidth]{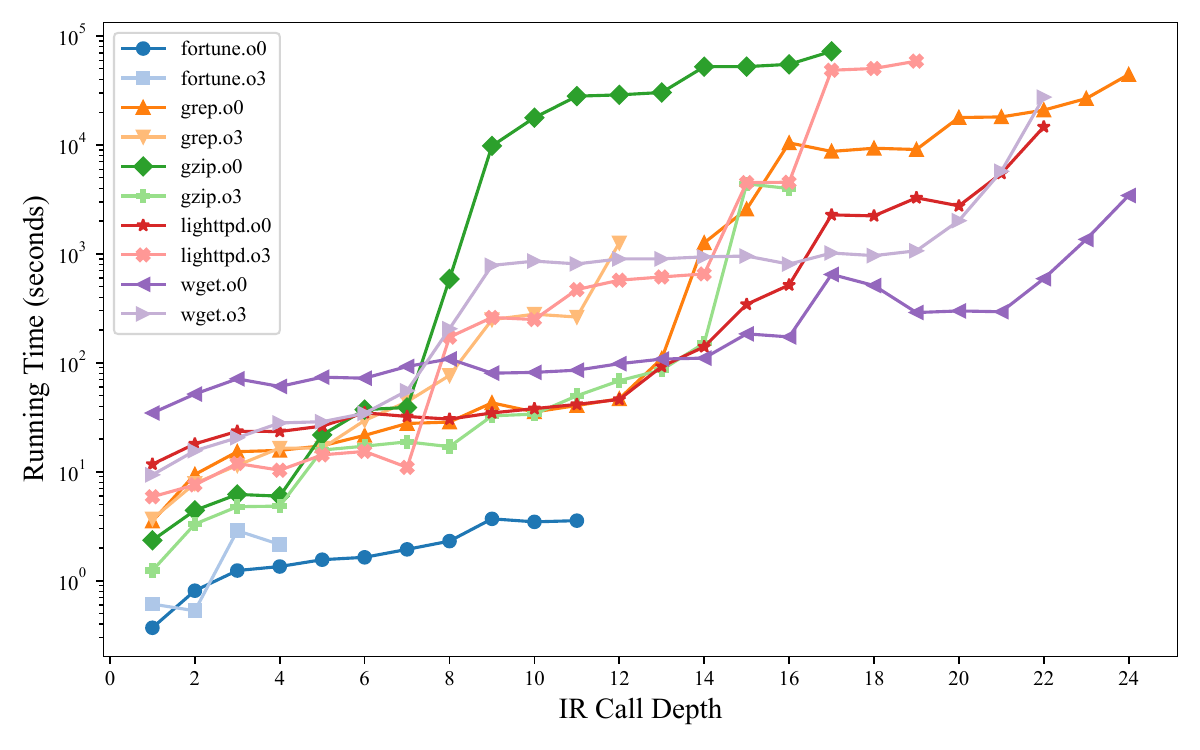}
    \caption{The time spent on inter-procedural type recovery analysis with the increase of IR call depth on the Howard dataset.}
    \label{rq1eff}
\end{figure}

To evaluate the efficiency, we decompile samples in both datasets using {\framework} and three baselines. 
We count the number of successful decompilations for each tool and also record the time and memory consumption. 
\revision{We set a 5-minute and 24-hour timeout for samples in Juliet and Howard dataset, respectively.}

\noindent \textbf{Results on Juliet}. As shown in \autoref{rq1-1}, we can observe that WaDec only successfully decompiles 1,940 samples. It gets stuck and does not output the final result for all the remaining 63.0\% (3,301) samples, reaching the timeout. After an investigation, we figure out that it is related to the phenomenon where large language models fall into endless repeated sequence output~\cite{llmloop}.

For memory and time consumption, Wasmdec consumes the least memory and time (6.46 MB, 0.02s) because it simply performs a conversion of WebAssembly instructions, lacking in-depth analysis. 
{\framework} consumes more memory and time (119.13 MB, 1.73 s) because of code optimization and type inference algorithms.
\revision{As for Ghidra, it consumes more memory and time (572.14 MB, 27.62s). This is because Ghidra exhibits a longer cold-start time due to its full-fledged disassembly and analysis framework written in Java that supports multiple architectures. In contrast, {\framework} is written in C++ and only contains decompiler and WebAssembly analysis logic, leading to its high efficiency.}
For WaDec, as it is based on a fine-tuned LLMs, the average peak memory usage is approximately 9.9 GB, while it also requires about 8.1 GB of GPU memory.

\noindent \textbf{Results on Howard}. 
\revision{The Howard dataset results are shown in \autoref{rq1-1-2}. As we can observe, with \texttt{-O3} enabled, the number of instructions in binaries decreases to 29\%--41\% of the original, leading to reduced time and memory consumption across all tools. {\framework} requires over 24 hours for six cases under unoptimized (\texttt{-O0}) conditions due to the high complexity of its adopted inter-procedural type recovery. 
To address this, we introduced ${\framework}_{F}$, which restricts type recovery to the intra-procedural level. For \texttt{-O3} cases, ${\framework}_{F}$ and Ghidra exhibit comparable decompilation times with ${\framework}_{F}$ consuming 43.77\% to 90.19\% less memory. 
With optimizations disabled (\texttt{-O0}), ${\framework}_{F}$ requires only a small fraction of Ghidra's time (ranging from 2.3\% on gzip to 22.7\% on lighttpd). We attribute this advantage to the compilation optimizations applied by {\framework} prior to decompilation (see \S~\ref{mid:lowpat}), which reduce the input size for subsequent code analysis. }
\revision{The results for WasmDec and WaDec remain consistent with previous observations. WasmDec, performing only simple conversions of WebAssembly instructions, continues to scale effectively to larger binaries. WaDec's decompilation time correlates with the scale of input and output and depends on GPU performance. However, we still observe occasional instances of endless repeated sequence output on certain functions, leading to unstable final decompilation times.}

\noindent \textbf{Complexity of the Type Recovery Algorithm}. \revision{To further depict and analyze the time complexity of the inter-procedural type inference algorithm employed by {\framework}, \textit{i.e.,} Retypd mentioned in \S~\ref{mid:retypd}, we examine the relationship between the time cost of type inference and call depth, as shown in \autoref{rq1eff}.
Since the y-axis is a logarithmic axis, we can easily observe the exponential relationship between them. Moreover, when reaching a particular depth, the analysis time even increases again dramatically (like \texttt{gzip-O0} from depth 7 to 9). 
This is because type summaries from deeper functions continuously accumulate summaries from callees. When a function calls multiple complex callees simultaneously, the constraint graph size expands several-fold, causing the original $O(n^3)$ algorithm to exhibit exponential time growth, eventually reaching the timeout we set in the experiment.}

\textit{\textbf{RQ1 Answer:} 
{\framework} demonstrates high efficiency on the Juliet dataset (1.73 s \& 119.13 MB) compared with Ghidra (27.62 s \& 572.14 MB). On the Howard dataset, while full {\framework} faces scalability challenges, its intra-procedural variant ${\framework}_{F}$ achieves a balance between time and decompilation capability, particularly showing advantages over Ghidra in processing unoptimized binaries, using less than one-quarter of Ghidra’s execution time and less than half its memory consumption.
}

\subsection{RQ2: Effectiveness} \label{ssec:rq2}

\revision{To comprehensively evaluate decompilation effectiveness, we assess 3 key aspects: recompilation success (syntactic correctness), Lines of Code (LoC, for conciseness), and type recovery accuracy. 
Specifically, we first measure recompilation success by compiling the decompiled C code with GCC. Next, we assess code conciseness through LoC measurement, following established practices in prior work~\cite{sailr,dream}.
Finally, we assess type recovery accuracy by counting the number of struct variable and struct member accesses recovered.}
Note that all baselines are not robust enough to be adopted directly. We discuss our fixes on them in \S~\ref{discussion}.

\begin{table}[t]
\centering
\caption{Recompilation success rate and average line of code on 5,241 samples from the Juliet dataset.}
\begin{tabular}{@{}r|l|l@{}}
\toprule
Tool        &  \# Recompile Success Rate  & Avg. LoC  \\ \midrule
{\framework}      & 5,241/5,241 (100.00\%)     & 87.60   \\
Ghidra      & 2,408/5,241 (45.95\%)     & 102.78  \\
WaDec       & 216/1,940 (11.13\%)      & 66.77   \\
Wasmdec     & 1,546/5,241 (29.50\%)    & 347.23   \\ \bottomrule
\end{tabular}
\label{rq2table}
\end{table}

\begin{table*}[t]
\centering
\caption{Recompilation success and line of code counts for the Howard dataset. {\framework} failed to decompile 6 programs within 24 hours, resulting in missing data.}
\resizebox{0.95\linewidth}{!}{

\begin{tabular}{@{}rlllllllllll@{}}
\toprule
\multirow{2}{*}{Benchmark} & \multicolumn{1}{c}{\multirow{2}{*}{\begin{tabular}[c]{@{}c@{}}\#Inst in \\ binary\end{tabular}}} & \multicolumn{5}{c}{Recompile Success} & \multicolumn{5}{c}{Line of Code} \\ \cmidrule(l){3-12} 
 & \multicolumn{1}{c}{} & {\framework} & \multicolumn{1}{c}{${\framework}_{F}$} & \multicolumn{1}{c}{Ghidra} & \multicolumn{1}{c}{WaDec} & \multicolumn{1}{c}{WasmDec} & \multicolumn{1}{c}{{\framework}} & \multicolumn{1}{c}{${\framework}_{F}$} & \multicolumn{1}{c}{Ghidra} & \multicolumn{1}{c}{WaDec} & \multicolumn{1}{c}{WasmDec} \\ \midrule
fortune (\texttt{O0}) & 13,653 & Y & Y & N & N & N & 2,404 & 2,306 & 1,982 & 890 & 8,914 \\
fortune (\texttt{O3}) & 4,797 & Y & Y & N & N & N & 2,135 & 2,084 & 2,017 & 87 & 1,822 \\
grep (\texttt{O0}) & 245,806 & Y & Y & N & N & N & 50,914 & 38,663 & 27,145 & 6,062 & 157,039 \\
grep (\texttt{O3}) & 93,227 & -- & Y & N & N & N & -- & 55,758 & 36,257 & 7,788 & 35,331 \\
gzip (\texttt{O0}) & 124,327 & -- & Y & N & N & N & -- & 17,575 & 11,040 & 2,295 & 79,624 \\
gzip (\texttt{O3}) & 51,279 & Y & Y & N & N & N & 28,574 & 25,521 & 19,157 & 4,762 & 16,630 \\
lighttpd (\texttt{O0}) & 484,209 & -- & Y & N & N & N & -- & 77,165 & 56,883 & 14,995 & 309,847 \\
lighttpd (\texttt{O3}) & 141,083 & -- & Y & N & N & N & -- & 80,115 & 54,995 & 40,119 & 48,677 \\
wget (\texttt{O0}) & 433,082 & -- & Y & N & N & N & -- & 63,911 & 51,866 & 10,252 & 276,468 \\
wget (\texttt{O3}) & 139,729 & -- & Y & N & N & N & -- & 77,854 & 56,629 & 21,284 & 54,189 \\
\bottomrule
\end{tabular}

}
\label{rq2table2}
\end{table*}

\noindent\textbf{Results on Juliet.} As shown in \autoref{rq2table}, {\framework} achieves a 100\% recompilation success rate on the Juliet dataset, demonstrating syntactic correctness in its decompiled results. 
WaDec achieves only 11.13\% recompilation success, primarily due to syntactic issues such as undeclared variables and missing global variables. WasmDec exhibits a 29.50\% recompilation success rate, with common syntax errors including incorrect insertion of break statements outside loop structures during WebAssembly-to-C conversion. 
Ghidra's recompilation success rate of 45.95\% is limited by its primary design focus on interactive reverse engineering rather than full-binary decompilation and recompilation workflows, exhibiting many syntax errors, such as conflicting types for global variables like ``\texttt{DAT\_ram\_0000093c}''.

For LoC metrics, WaDec produces the lowest average line count (66.77). However, further investigation reveals frequent statement omissions and the presence of operations not present in the original code, which may be attributed to the common hallucination issues associated with large language model. 
{\framework} achieves a much lower average LoC (87.60) compared to Ghidra (102.78). This reduction demonstrates the effectiveness of the in-depth type analysis described in \S~\ref{method:tr}, which minimizes stack variables and transforms complex pointer operations into struct member accesses, thereby enhancing both code conciseness and readability. WasmDec, on the other hand, yields the highest average LoC (347.23) as it fails to convert low-level operations (such as stack pointer subtraction) into equivalent C semantics, resulting in significantly less readable output.

\noindent\textbf{Results on Howard.} 
\revision{As shown in \autoref{rq2table2}, {\framework} and ${\framework}_{F}$ are the only tools that consistently achieved successful recompilation across all five programs in the Howard dataset (for successfully decompiled cases). In contrast, all other baselines consistently fail to recompile the decompiled code, exhibited persistent syntactic errors, similar to those observed in the Juliet dataset. Regarding code conciseness, ${\framework}_{F}$ produces higher LoC than Ghidra, which attributes to the declaration of recovered custom structures, which can be up to tens of thousands LoCs, like 19K for \texttt{lighttpd-O3}. WasmDec's output LoC directly correlates with the binary instruction count, yielding significantly more lines for unoptimized (\texttt{-O0}) binaries compared to optimized (\texttt{-O3}) versions.
WaDec, while producing the lowest LoC count, frequently failed on larger functions due to "input too long" errors in LLMs, likely because it exceeded the model's maximum context window capacity.}

\noindent\textbf{Type Recovery Accuracy.}
\label{sssec:structrec}
\revision{To further evaluate the type recovery accuracy (detailed in \S~\ref{sec:imp}), we manually inspected 319 functions (269 functions from randomly selected 50 Juliet cases, and 5 randomly selected functions from each of the 5 Howard programs at both \texttt{-O0} and \texttt{-O3} optimization levels). For a complete comparison, we use results of ${\framework}_{F}$ for the Howard dataset. We obtained the original source code for each function and compared it with the decompiled results.}

\revision{
On the Juliet dataset, functions are relatively simple. Among the 269 sampled functions, we identified only 6 struct-typed variables. Our approach successfully reconstructed all of them as proper struct types, whereas Ghidra type their pointers as plain integers. Furthermore, we correctly converted all 14 struct member accesses in Juliet into struct field accesses. Ghidra, in contrast, retained them as raw pointer arithmetic and dereference operations.}

\begin{figure}[t]
    \centering
    \includegraphics[width=\linewidth]{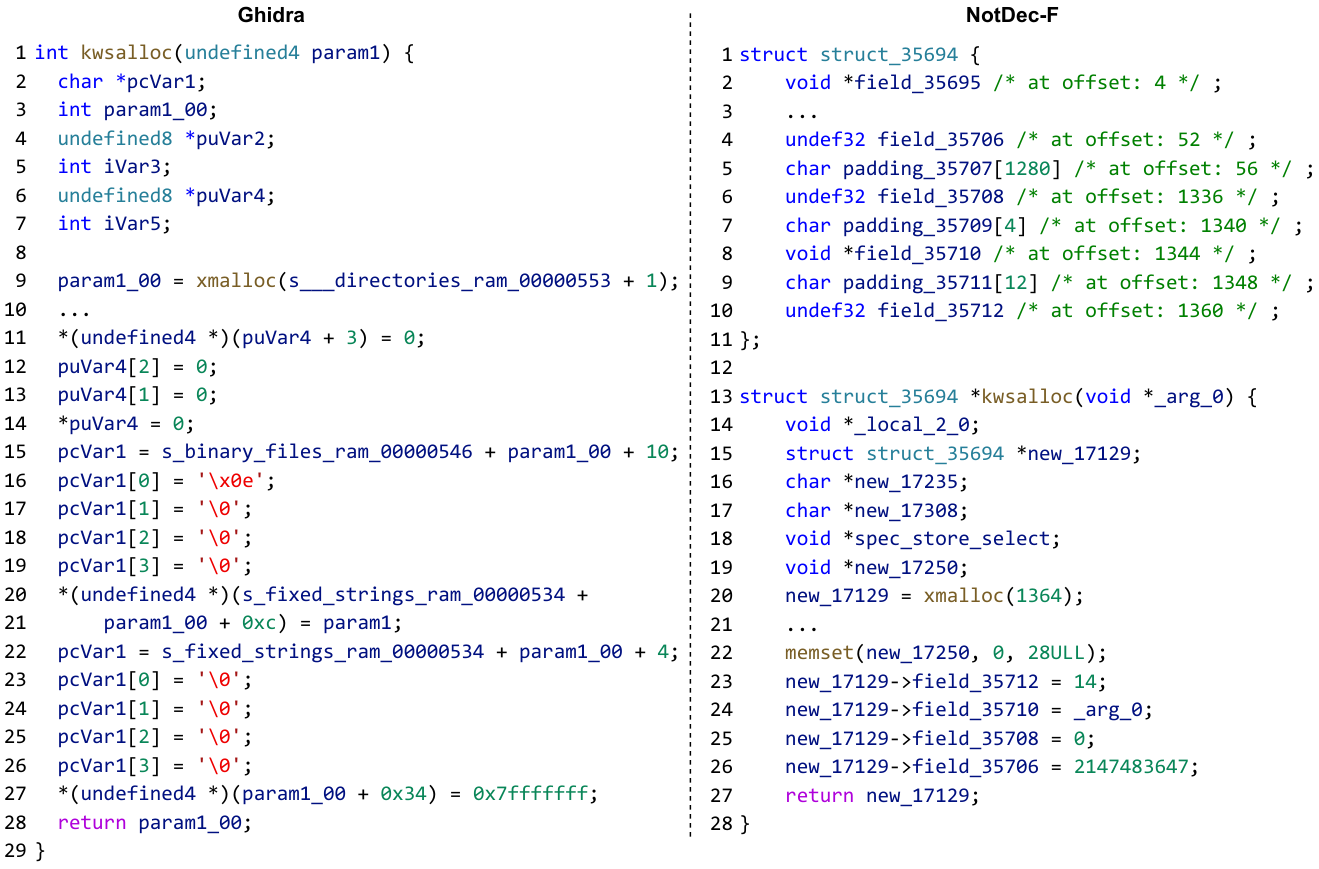}
    \caption{Decompilation results of \texttt{kwsalloc} from \texttt{grep}, where {\framework} can recover fields of the custom structure used.}
    \label{rq2sample2}
\end{figure}

\revision{For the Howard dataset, there are 29 struct definitions and declarations that are accessed. ${\framework}_{F}$ successfully reconstructed 24 (82.76\%) of them. For the remaining five, we found that four of them lack any member accesses within the current function, and one global struct is flattened during decompilation, with its members treated as independent global variables. 
In contrast, Ghidra identified only 4 (13.79\%) of the 29 structs, primarily recognizing standard types like \texttt{timespec} and \texttt{stat} through applying known library function signatures.
Regarding struct member accesses, the source code contains 368 such accesses. ${\framework}_{F}$ correctly recovered 314 (85.33\%) of them. The remaining 54 accesses correspond to members of global structs that were flattened during decompilation, which, consequently, were directly mapped to the deconstructed global variables. 
Ghidra, on the other hand, recovered only 34 (9.24\%) member accesses. Beyond the same 54 flattened global accesses, an additional 241 were left as raw pointer arithmetic and dereferences, and 39 were represented as array-indexed accesses instead of struct field references.
Take the \texttt{kwsalloc} function from \texttt{grep} as an example, shown in \autoref{rq2sample2}. ${\framework}_{F}$ correctly reconstructed custom structure allocation at line 20, and converted those structure member accesses at line 23-26, while Ghidra could only represent these operations as low-level pointer arithmetics and array assignments at line 11-27.}

\textit{\textbf{RQ2 Answer:} 
{\framework} achieves a 100\% recompilation success rate, far exceeding Ghidra's 45.95\% on the Juliet dataset. In terms of type recovery accuracy, the variant ${\framework}_{F}$ can recover 82.76\% structure variables and 85.33\% struct member accesses while Ghidra can only recover 13.79\% structure and 9.24\% member accesses on samples from the Howard dataset.}

\subsection{RQ3: Ablation Study} \label{eva:abla}

Regarding {\framework}, we underline that only the optimization and type recovery module (see \S~\ref{method:tr}) can be removed without affecting the functional integrity. Removing the code lifting module (see \S~\ref{method:lift}) prevents the tool from parsing WebAssembly code, while removing the C code generation module (see \S~\ref{method:cgen}) results in the tool generating only an intermediate representation, making it unable to produce C language code.

To study the impact of different levels of code optimization and type recovery on decompilation results, we compare {\framework} with the following three variants:
\begin{itemize}[leftmargin=*]
\item ${\framework}_{d0}$: without any code optimization and type recovery.
\item ${\framework}_{d1}$: only running the code optimization and low-level pattern matting (\S~\ref{mid:lowpat}).
\item ${\framework}_{d2}$: performing type recovery while treating the stack as a whole structure.
\end{itemize}
We conducted the same experiment under the same settings of RQ1 and RQ2. Due to scalability of our type recovery algorithm, we perform ablation study only on the Juliet dataset.

\begin{table}[t]
\centering
\caption{Ablation study results in terms of efficiency, syntactic correctness, and code conciseness.}
\resizebox{\linewidth}{!}{
\begin{tabular}{@{}rccccc@{}}
\toprule
Tool & \# Success & \begin{tabular}[c]{@{}c@{}}Avg.\\Time (s)\end{tabular} & \begin{tabular}[c]{@{}c@{}}Avg.\\Memory (MB)\end{tabular} & \begin{tabular}[c]{@{}c@{}}\# Recompile\\ Success\end{tabular} & Avg. LoC \\ \midrule
${\framework}_{d0}$ & 5,241 & 0.14 & 85.16 & 5,241 & 455.36 \\
${\framework}_{d1}$ & 5,241 & 0.94 & 87.83 & 5,241 & 131.62 \\
${\framework}_{d2}$ & 5,241 & 1.71 & 110.45 & 5,241 & 131.22 \\
{\framework} & 5,241 & 1.73 & 119.13 & 5,241 & 87.60 \\ \bottomrule
\end{tabular}
}
\label{rq3table}
\end{table}

\noindent\textbf{Efficiency.}
In \autoref{rq3table}, all variants can successfully decompile all samples. For efficiency, the average decompilation time gradually increases with the introduction of optimizations and type recovery. The overhead of ${\framework}_{d0}$ is the lowest, at only 0.14s, followed by ${\framework}_{d1}$ (0.94s), while ${\framework}_{d2}$ and {\framework} have similar times of 1.71s and 1.73s, respectively. Memory consumption shows a similar trend: ${\framework}_{d0}$ has the least memory usage (85.16 MB), while the complete type recovery {\framework} uses the most (119.13 MB). This shows that type recovery, especially advanced analyses such as structure recognition, increases resource overhead.

\noindent\textbf{Effectiveness.}
As in RQ2, we assess the variants from three aspects: syntactic correctness (via recompilability), code conciseness (by LoC), and type recovery accuracy. 
For syntactic correctness, all variants produce recompilable code for all 5,241 samples. In terms of code conciseness, the code generated by ${\framework}_{d0}$ is the most verbose, averaging 455.36 LoC. 
The introduction of compilation optimizations in ${\framework}_{d1}$ effectively eliminates redundant codes, significantly reducing the average LoC to 131.62. 
${\framework}_{d2}$ further simplifies the code through type analysis. However, by treating the entire function stack as a structure, it inserts many structure definitions, resulting in an average LoC similar to ${\framework}_{d1}$ (131.22). The fully configured ${\framework}$ produces the most concise code, with an average of only 87.6 LoC, by splitting the stack. Overall, as optimization and type recovery strategies are gradually introduced, the conciseness of the decompiled code continue to improve.

\revision{For type recovery accuracy, similar to RQ2, we count the number of recovered structure variables and member accesses on the sampled 269 functions, as shown in \autoref{rq3table2}. Since the underlying WebAssembly instructions contain no structural information, ${\framework}_{d0}$ recovers no structure variables or member accesses. ${\framework}_{d1}$, which only applies additional code optimizations, exhibits similar behaviors. 
By interpreting the entire stack as a structure, ${\framework}_{d2}$ creates a substantial number of structure variables (275) and member accesses (1,905). In contrast, ${\framework}$ further refines this approach by splitting the stack structure, converting most member accesses into direct variable accesses. Manual inspection of the source code for these 269 functions confirms that all recovered structure variables and member accesses are accurate and complete, with no omissions.}

\begin{table}[t]
\centering
\caption{Ablation study results in terms of type recovery accuracy on sampled 269 functions of Juliet dataset.}
\begin{tabular}{@{}rcc@{}}
\toprule
Tool & \begin{tabular}[c]{@{}c@{}}\# Recovered\\Structure Variable\end{tabular} & \begin{tabular}[c]{@{}c@{}}\# Recovered\\Member Access\end{tabular} \\ \midrule
${\framework}_{d0}$ & 0 & 0 \\
${\framework}_{d1}$ & 0 & 0 \\
${\framework}_{d2}$ & 275 & 1,905 \\
{\framework} & 6 & 14 \\ \bottomrule
\end{tabular}
\label{rq3table2}
\end{table}

\textit{\textbf{RQ3 Answer:} Through an ablation study, we found that code optimization combined with low-level pattern matching effectively reduces redundancy in binary code, achieving about 70\% reduction in lines of code (LoC). The type recovery module contributes to the recovery of 100\% of struct types and their associated accesses. Furthermore, deconstructing the stack into concrete variables yields an additional 30\% reduction in LoC.}

%% file: discussion.tex
\section{Related work}

\noindent
\textbf{WebAssembly Analysis Techniques}. 
WebAssembly analysis resembles traditional binary analysis. 
Wassail\cite{wassail} applies a compositional analysis technique to find vulnerabilities in a binary. 
Wasmati~\cite{wasmati} employs Code Property Graphs (CPG) with pattern matching to find vulnerabilities. MINOS~\cite{minos} detects cryptocurrency-related operations by extracting runtime features like instruction-type distributions. Eunomia~\cite{eunomia} applies symbolic execution to WebAssembly code.
However, Due to the complexity of analyzing linear memory, these approaches often face high complexity (e.g., symbolic execution's path explosion) or reduced accuracy.

\noindent
\textbf{Native Decompilation}. The conversion of control flow graphs into control flow statements has been thoroughly resear\-ched~\cite{dream,phoenix,comb,sailr}. Recently, there has been an increasing number of works on decompiling binary code based on large language models~\cite{llmdec1,llmdec2}. However, methods based on large models face issues of hallucination, and the results lack interpretability.

\noindent
\textbf{Binary Type Recovery}
For type recovery, previous type inference schemes~\cite{tie,secondwrite} often relied on the results of pointer analysis. Retypd~\cite{retypd} proposed an inter-procedural polymorphic type recovery scheme that does not require pointer analysis. There is also an attempt to simplify the Retypd algorithm~\cite{binsub}. In recent research, Osprey attempted to recover types in binary code based on probabilistic sampling~\cite{osprey}. BinPointer proposed an accurate pointer analysis at the binary level~\cite{binpointer}. Manta~\cite{manta} introduced hybrid-sensitive type analysis and implemented binary bug detection. 

\section{Discussion} \label{discussion}
\noindent
\textbf{Effectiveness Threats}. 
Four factors may affect effectiveness.
\one \revision{Our type recovery approach does not support indirect calls, which compromises the accuracy of type inference. This limitation can be addressed by incorporating function pointer analysis to resolve the targets of indirect calls.}  
\two \revision{Our approach relies on assumptions for recognizing low-level features, which, if unmet, threaten the effectiveness. These include the heuristic for stack allocation failing if the global stack pointer is not found, and pattern matching for libc functions (e.g., memset, memcpy) failing when they are inlined by LTO. Similarly, the heuristic for detecting variadic functions by matching argument arrays may be incomplete, requiring manual pattern updates.} 
\three If PNDiff analysis does not have enough information to recognize all pointer arithmetic, it will result in the absence of hints for some structure members. This can be solved by allowing users to provide more information about which variable is pointer or number. 
\four The Memory SSA used in the backend code generation process may face over-approximation issues. However, we use Memory SSA solely to determine whether to insert temporary variables to hold the results of memory loads. In cases of uncertainty, our approach conservatively assumes potential aliasing and inserts the temporary variable, thereby avoiding any incorrect omission that could alter program semantics.

\noindent
\textbf{Efficiency of Type Recovery}. 
\revision{
Type recovery for WebAssembly faces challenges similar to those encountered in native binaries.
Specifically, the Retypd type recovery algorithm we employed is not sufficiently fast. As shown in \S\ref{ssec:rq1}, inter-procedural analysis must be disabled to scale to large binaries. We underline that this inefficiency arises partly because Retypd implicitly performs computations analogous to those in pointer analysis. Explicitly decoupling and exposing pointer analysis results could therefore enable a broader range of downstream analyses that rely on points-to information. Furthermore, binary type recovery is fundamentally a form of type inference, a problem extensively studied in programming language (PL) theory. Recent advances in PL-type inference, particularly efficient algorithms supporting polymorphism and subtyping~\cite{simplesub,mlsub}, offer promising pathways toward more accurate and scalable type recovery techniques for binary code.
}

\noindent
\textbf{Choice of Decompiler IR.} \revision{We opted to use LLVM IR as the intermediate representation for decompilation. This decision is supported by the fact that existing decompilers already adopt SSA-based design principles and optimization algorithms, such as Ghidra's P-Code and Binary Ninja's IL. Since semi-SSA forms are capable of representing both SSA and non-SSA code, they exhibit no fundamental divergence from LLVM IR in practice. Furthermore, we aim to faithfully capture the complete semantics of the binary in the IR, potentially through custom intrinsic functions or externally maintained mappings.}

\noindent
\textbf{Other Source Programming Languages}. We underline that decompiling the WebAssembly code compiled by other languages can largely follow the method proposed in this paper. 
Emerging languages such as Go and Rust also adopt a stack and heap-based memory allocation model. To support these languages, it is necessary to write targeted patterns for stack and heap memory allocation matching (\S\ref{mid:lowpat}), as well as to develop a dedicated backend to generate the corresponding syntax tree for them (\S\ref{method:cgen}). 

\noindent
\textbf{Obfuscated WebAssembly Code}. We assume that the input of {\framework} is generated by a common compiler. Though {\framework} can handle simple arithmetic obfuscation that can be recognized and reverted by the compiler's optimization, other complex code obfuscations can significantly negatively impact the results. 
For example, obfuscation of stack pointer operations can affect the recognition of stack memory allocation, thereby impacting the identification of local variables.
We leave this as a future direction.

\noindent
\textbf{Fixes on Baselines.} We applied lightweight patches to each decompiler to eliminate obvious syntax errors: for Ghidra (v10.3.2), we merged the pending upstream commit\footnote{https://github.com/NationalSecurityAgency/ghidra/commit/7559acf5} to enable global‐variable exports, then used simple pattern matching to correct string globals from the unsupported \texttt{string} type to valid C types (\texttt{char*}); for \texttt{wasmdec}, we initialized an uninitialized class flag to zero (preventing garbage‐dependent missing semicolons) and resolved type‐mismatch errors in load/store instructions; and for WaDec, \texttt{wasmdec}, and Ghidra outputs, we declared all called functions upfront to satisfy inter‐function dependencies, additionally fixing WaDec’s malformed \texttt{main} signature via a straightforward regex replacement.

\section{Conclusion}
This paper presents a complete WebAssembly‐to‐C decompilation pipeline {\framework}, organized into three phases: \one 
Code Lifting: parses Wasm bytecode and converts it into an SSA‐based IR, exposing use‐def chains. \two Optimization \& Type Recovery: applies compiler optimizations and Retypd enhanced with pointer and numeric type differentiation analysis (PNDiff) to recover high‑level C types, and splits stack with modified scalar replacement of aggregate (SROA) optimization.
\three Control‑Flow Structuring \& C Code Generation: uses Memory SSA for correct and readable transformation of non-control-flow instructions, restructures high‐level control constructs with semantic-preserving algorithm, and emits readable C code.
Evaluation on 5,241 Juliet samples and five real-world programs demonstrates that {\framework} is the only tool achieving 100\% recompilation success across all binaries, while recovering 85.33\% of struct member accesses compared to Ghidra's 9.24\%, with 15.9× faster execution and 4.8× lower memory consumption.